# Predicting mode-locked fiber laser output using feed-forward neural network

XINYANG LIU [1,*] AND REGINA GUMENYUK [1,2]

[1]*Laboratory of photonics, Tampere University, Korkeakoulunkatu 3, Tampere 33720, Finland*
[2]*Tampere Institute for Advanced Study, Tampere University, Kalevantie 4, Tampere 33100, Finland*
*\*xinyang.liu@tuni.fi*

**Abstract:** With a great ability to solve regression problems, the artificial neural network has become a powerful tool to facilitate advancing ultrafast laser research. In this contribution, we demonstrate the capability of a feed-forward neural network (FNN) to predict the output parameters of a mode-locked fiber laser, which mutually depend on multiple intracavity parameters, with high speed and accuracy. A direct mapping between cavity parameters and laser output is realized through the FNN-trained models, bypassing tedious iterative numerical simulation as a common approach to get a converged solution for a laser cavity. We show that the laser output spectrum and temporal pulse profiles can be accurately predicted with the normalized root mean square error (NRMSE) of less than 0.032 within only a 5 ms time frame for scenarios inside and outside the training data. We investigate the influence of FNN configuration on prediction performance. Both gain and spectral filter parameters are explored to test the prediction capability of the trained FNN models at high speed. Straightforward and fast prediction of the laser output performance for varying laser intracavity parameters paves the way to intelligent short-pulsed lasers with inversed design or autonomous operation maintenance.

## 1. Introduction

In recent years, machine learning, which is well-known for its artificial intelligence brought by biology-inspired algorithms, has been coupled with ultrafast photonics research, dramatically advancing the field [1]. Problems relating to regression and optimization can be solved more effectively and efficiently than conventional methods. The optimization problem solver, represented by the natural-selection-inspired genetic evolutionary algorithm, is successfully employed for automatic searching/switching of specific laser operation state [2-4], tailoring of supercontinuum [5] and flat frequency comb generation [6], revealing the versatility of the single algorithm. The regression problems are commonly solved by artificial neural network, which is inspired by brain structure. In laser pulse characterization, single-shot fast retrieval of both laser pulse amplitude and phase has been realized [7-9], as well as features of chaotic and unstable laser pulses [10,11]. The spatial mode composition is also predicted by neural networks [12]. In pulse shaping, a neural network is used to build a physics-informed model to prescribe input conditions for desired output or inversely predict the parameters of the pulse shaping system in both the time domain and spatial domain [13-15]. Inside the fiber laser cavity, the ability of the neural network to harness the pulse evolution dynamic is also demonstrated [16]. Conversely, to increase the computation speed and energy consumption efficiency of neural networks, optical systems are employed to build optical neural networks [17-19].

As a main workforce in photonics research, ultrafast laser pulses are dominantly generated from laser resonators, including solid-state laser cavities with rear-earth-ion doped crystal/glass or optical fibers and semiconductors as gain material. Ultrafast laser pulses with a pulse duration of a few tens fs to a few ps are readily generated and commercially available. As indicated by the master equation for mode-locking [20], laser pulse formation within the laser cavity requires a complex balance of nonlinearity/dispersion and gain/loss. Thus, the output laser pulse properties, including pulse duration, pulse energy, spectral width and phase profile, mutually depend on multiple cavity parameters. In anomalous-dispersion fiber laser cavities, the configuration of chromatic dispersion balances phase modulation, which, with the appropriate combination of gain and loss, leads to a stable laser pulse formation. In normal-dispersion fiber laser cavities, the dissipative process comes into play. A balance between dissipative effects, such as spectral filtering and intracavity spectrum broadening, is additionally required for pulse formation.

Fiber laser is commonly simulated by the generalized nonlinear Schrödinger equation (GNLSE), with different-valued physical parameters for different parts of the laser cavity. The simulation typically starts with white noise. After many round trips of iterative propagation, the white noise can converge to a stable self-consistent pulse if the cavity parameters are appropriately set. In a fiber laser cavity, chromatic dispersion and Kerr nonlinearity are coupled with each other through optical fiber, meaning neither of them can be independently tailored since the variation of fiber length would change both parameters simultaneously. Especially, the phase modulation contributed by Kerr nonlinearity is also linked with laser intensity, and laser intensity, in turn, is affected by gain/loss parameters. This makes the cavity parameters interdependent. In practical research, if one needs to model laser dynamics through numerical simulation, a tedious and time-consuming process of sweeping cavity parameters is indispensable. The neural network has been recently utilized as an approximator to directly map the output parameters of mode-locked fiber lasers [8, 16, 21, 22], bypassing the repetitive step-by-step time-consuming computation. Deep, recurrent or artificial neural networks demonstrated the powerful capability to predict the spectral and/or temporal pulse profiles within several hundreds of milliseconds to a few seconds time frame, which is orders of magnitude shorter than computational time for the GNLSE-based approach. In this study, we investigate the capability of the FNN-trained model to approximate the output performance of a mode-locked fiber laser with excellent accuracy (NRSME is below 0.032) and as fast as 5 ms computational time. We train different FNN model configurations with varying numbers of hidden layers and neurons to evaluate their impact on prediction accuracy. To examine the FNN-trained models, we chose gain and spectral filter bandwidth to be variable cavity parameters. Variation of both parameters triggers changes in several other intracavity parameters, therefore performing the multi-variable complex cases. We demonstrate that using the trained FNN models, both temporal and spectral intensity profiles of laser output can be accurately predicted at high speed.

## 2. Configuration of numerical simulations

A linear fiber laser cavity, which mimics a practical cavity, is employed as a test platform. The schematic illustration is shown in Fig. 1 (a). The cavity is terminated by a reflective mirror (M) and a saturable absorption mirror (SAM). L1-L4 indicates four pieces of passive fiber, with a length of 25 cm each. 50 cm gain fiber is placed between L2 and L3, providing optical gain centered at 1040 nm with a 45 nm gain bandwidth. A gaussian-shaped spectral filter (G-filter) with a central wavelength of 1040 nm is placed between L1 and L2. An output coupler (OC) is placed between L3 and L4 and couples out 10% of optical power. The total cavity length is 1.5 m corresponding to ~33 MHz repetition rate.

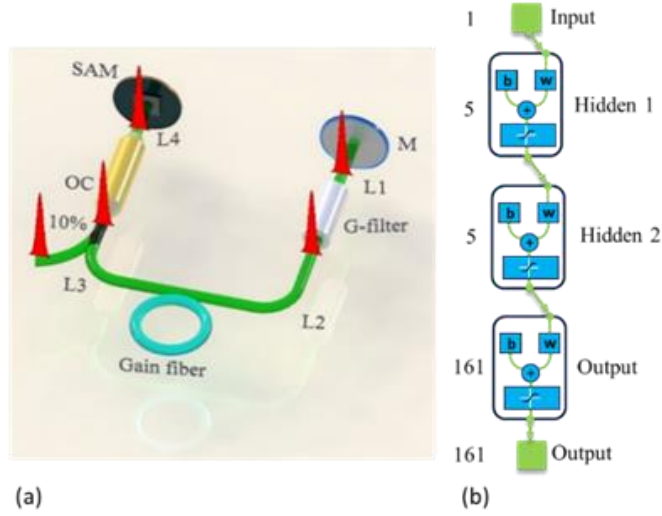

Fig. 1. (a) Sketch of the linear fiber laser cavity. (b) FNN model structure with 2 hidden layers of 5 neurons.

First, we generate the training data by the numerical simulation of laser field propagation within the laser cavity, going through different parts one by one. Fiber part is simulated by the generalized nonlinear Schrödinger Equation (GNLSE):

$$\frac{\partial A}{\partial z} + i\frac{\beta_2}{2}\frac{\partial^2 A}{\partial T^2} = \frac{g}{2}A + i\gamma\left(1 + i\tau_{shock}\frac{\partial}{\partial T}\right)\left(A\int_{-\infty}^{\infty} R(t)|A(z,T-t)|^2 dt\right). \quad (1)$$

where A(z,T) is the complex electrical field, z is the coordinate of pulse propagation, T is the retarded time coordinate moving at the group velocity of pulse, $\beta_2$ is the group velocity dispersion (GVD) with a value of 0.021 ps²/m, g is the gain coefficient for gain fiber, γ is the nonlinear coefficient with a value of 0.0041 W⁻¹m⁻¹, $\tau_{shock}$ is the shock-formation time-scale that is responsible for the self-steeping effect and R(t) is the response function that models both the instantaneous electronic (Kerr) nonlinearity and the delayed molecular (Raman) nonlinearity [23]. Equation (1) is solved by the split-step Fourier method using the second order Runge-Kutta algorithm. The gain coefficient provided by a gain fiber is expressed by the saturable model:

$$g = \frac{g_0}{1+\frac{E_{pulse}}{E_{sat}}}. \quad (2)$$

where $g_0$ is the small-signal gain coefficient with values following gain profile, $E_{pulse}$ is the pulse energy with a value calculated by integrating the optical power within the simulated time window and $E_{sat}$ is the gain saturation energy. The gain profile is implemented in the spectral domain with a Gaussian shape and a full width at half maximum (FWHM) of 45 nm. In passive fiber, the gain coefficient is set to 0.

The SAM is modelled by the following saturable reflectance equation:

$$R = R_s + dR\left(1 - \frac{1}{1+\frac{P}{P_{sat}}}\right). \quad (3)$$

where R is the reflectance of SAM, $R_s$ is small-signal reflectance with a value of 0.42, $dR$ is modulation depth of SAM with a value of 0.5, $P_{sat}$ is saturation power with a value of 16.9 W and P is instantaneous power of laser field. The saturated loss is 0.08.

The time window for the simulation is 50 ps with 1024 data points. The small signal gain (SSG) for gain fiber and filter bandwidth (FBW) for G-filter are chosen as variable cavity parameters, separately. The laser cavity simulation starts with white noise. After less than 350 roundtrips iteration for all variable parameter values, the laser field can converge to a self-consistent single-pulse solution. The maximum roundtrip number is fixed to 500 in simulations. The output intensity profiles are normalized since we only focus on the intensity profile prediction in this study. To prepare the data for training FNN models and increase training speed, 1024 data points are trimmed to 161 data points while preserving the converged pulse profiles.

The FNN models are trained in MATLAB R2020b using the built-in functions. The employed FNN models have one input layer, variable hidden layers of variable neurons and one output layer. Fig. 1 (b) illustrates the structure of a FNN model with 2 hidden layers of 5 neurons as an example. The input layer has one neuron since only one cavity parameter with variable values is set as the input vector for the input layer. To study the influence of the number of hidden layers and neurons on FNN model prediction accuracy, FNN models with 1-3 hidden layers of 1-5 neurons are trained separately in a supervised learning manner. The output layer has 161 neurons, determined by the size of the output vector for the output layer. For both temporal and spectral domains intensity profile predictions, 15 FNN models are trained for each domain with different hidden layers and neurons configurations.

## 3. Results and discussion

First, SSG is set as a variable cavity parameter. The SSG value is varied from 2 m-1 to 22 m-1 with a step of 0.01 m-1. To train the FNN model, an array of SSG with values from 2 to 20 with a step of 0.01 works as the input vector for the input layer, and the corresponding laser cavity outputs of temporal/spectral intensity profiles work as the output vector for the output layer. SSG values of 20.01 m-1 to 22 m-1 are excluded to test the prediction capability of the trained FNN models for the dataset that is outside training data. SSG of 11 m-1 and 22 m-1 are chosen to evaluate the prediction performance. The normalized root mean squared error (NRMSE) calculated by following equation is used to measure the prediction error.

$$NRMSE = \sqrt{\frac{\sum(x_F - x_G)^2}{\sum x_G^2}} \qquad (4)$$

where $x_F$ is the FNN model predicted data and $x_G$ is the GNLSE simulated data.

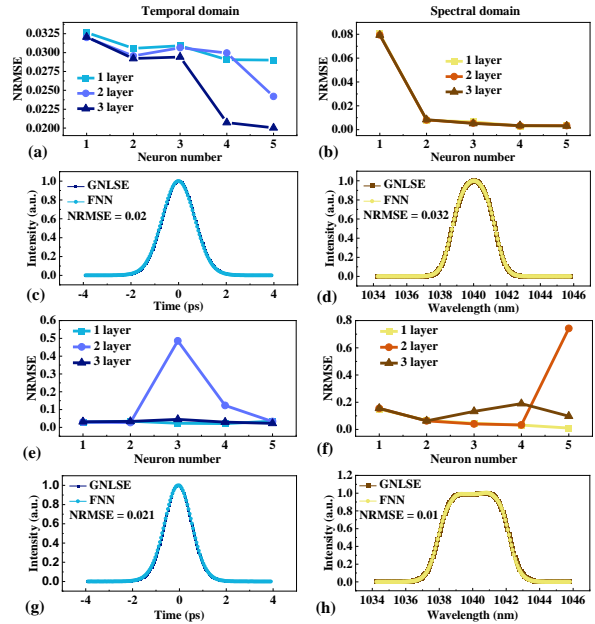

Fig. 2. NRMSEs of FNN model prediction for SSG of 11 m-1 (a, b) and 22 m-1 (e, f) in temporal and spectral domains. (c, d, g, h) Comparison between FNN model predicted and GNLSE simulated intensity profile in temporal and spectral domains, respectively, using the FNN model with lowest NRMSE in each case.

NRMSEs of FNN model prediction for SSG = 11 m-1 are summarized in Fig. 2(a) and Fig. 2(b) for temporal and spectral domains, respectively. For temporal domain prediction, all NRMSEs are below 0.033 for different hidden layers and neuron configurations. Although there is no clear dependency of prediction accuracy on the number of hidden layers and neurons, the prediction performance is improved when the neuron number is larger than 1. For spectral domain prediction, the NRMSEs decrease from ~0.08 to below 0.01 when the neuron number is larger than 1. However, the number of hidden layers does not clearly affect prediction performance. Fig. 2(c) and Fig. 2(d) illustrate the comparison between GNLSE simulated and FNN model predicted temporal/spectral intensity profiles. The FNN models used for Fig. 2(c) and Fig. 2(d) are the ones with the lowest NRMSE, i.e., the FNN models with 3 hidden layers of 5 neurons for both temporal and spectral domain intensity profile prediction. A good match between the GNLSE simulated profile and the FNN model predicted profile is realized. To test the prediction capability of the trained FNN models for a cavity parameter outside the training dataset, an SSG of 22 m-1 is chosen. Fig. 2(e) and Fig. 2(f) summarize NRMSEs for FNN model prediction in the temporal/spectral domain with different numbers of hidden layers and neurons. For temporal domain prediction, the FNN models with 1 hidden layer and 2 hidden layers operate with high accuracy, and corresponding NRMSEs ranging from 0.023 to 0.045. While the FNN model with 3 hidden layers has scattered NRMSEs from 0.027 to 0.486. For spectral domain prediction, the NRMSEs scatter from 0.01 to 0.742. Compared with the trained case of SSG = 11 m-1 (Fig. 2(a) and Fig. 2(b)), the prediction performance of FNN models in the case outside the training data of SSG = 22 m-1 demonstrates a wider range of errors and random dependency on layers and neuron numbers. The increase in NRMSE range could be associated with a long cavity length, corresponding to a 1.5 km field propagation length for 500 roundtrips number. Long-range propagation is a highly challenging scenario for neural network prediction, as a mode-locked laser cavity incorporates loss, gain, dispersion, nonlinearities, saturable absorption and other effects [22]. However, the FNN models can still predict the laser output with NRSME below 0.03 for cavity parameters outside the training dataset, as shown in Fig. 2(g) and Fig. 2(h). The above results confirm that well-trained FNN models can be utilized to do the direct mapping from cavity parameters to the equilibrium solution of the laser cavity, bypassing the many round-trip iterations of the laser field in conventional simulation.

To test the universality of FNN model for direct mapping laser output, the FBW of G-filer is then set as variable cavity parameter. The FBW is varied from 3 nm to 13 nm with a step of 0.01 nm. The array of 3 to 11 with a step of 0.01 is set as the input vector for the input layer at the training stage of FNN models and the corresponding laser outputs of temporal/spectral intensity profiles are set as the output vector for the output layer. FBW of 11.01 nm to 13 nm are excluded for testing the prediction capability of the trained FNN models for data outside training dataset. FBW of 7 nm (trained data parameter) and 13 nm (outside of trained data parameter) are chosen to evaluate the prediction performance of trained FNN models. In case of FBW = 7 nm, Fig. 3(a) and Fig. 3(b) depict the NRMSEs for FNN model predictions in temporal domain and spectral domain, respectively. For temporal domain predictions, the NRMSEs vary from 0.007 nm to 0.043 nm for different hidden layers and neurons configurations. For spectral domain predictions, the NRMSEs change from 0.0006 nm to 0.042 nm. No clear dependency of NRMSEs on amounts of hidden layers and neurons can be identified. Fig. 3(c) and Fig. 3(d) compare the GNLSE simulated profiles and the FNN model generated profiles. The FNN models applied on Fig. 3(c) and Fig. 3(d) are the ones with lowest NRMSEs, i.e., the FNN models with 3 hidden layers of 5 neurons and 1 hidden layer of 5 neurons, respectively. Negligible discrepancy confirms the good prediction accuracy of FNN models. FBW of 13 nm is used to test the FNN model prediction capability for data outside training dataset. NRMSEs are summarized in Fig. 3(e) and Fig. 3(f) for temporal and spectral domains, respectively. The NRMSEs of temporal domain prediction scattered from 0.024 to 0.224, while for spectral domain prediction, NRMSEs change from 0.0397 to 0.243. Despite the increased values of

NRMSEs compared with the case of FBW = 7 nm, the GNLSE simulated profile and the FNN model generated profile (Fig. 3(g) and Fig. 3(h)) still show great match. The FNN models applied in Fig. 3(g) and Fig. 3(h) are the ones with 1 hidden layer of 4 neurons and 3 hidden layers of 3 neurons, respectively. The good prediction accuracy when using FBW as a variable cavity parameter confirms the universality of FNN models as a direct mapping tool to predict laser output.

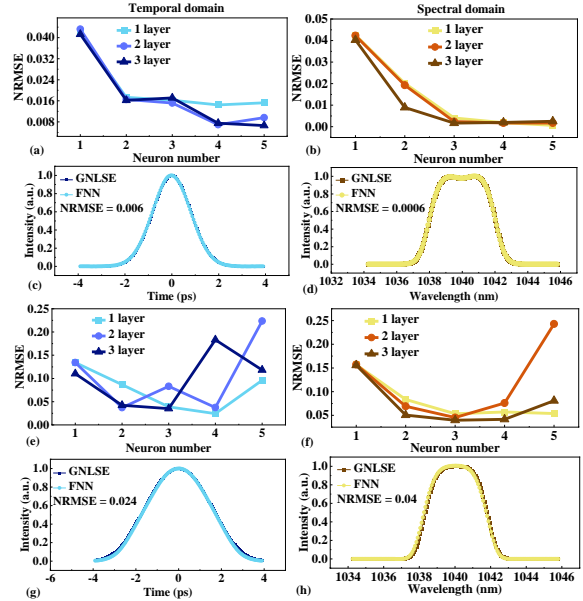

Fig. 3. NRMSEs of FNN model prediction for FBW of 7 nm (a, b) and 13 nm (e, f) in temporal and spectral domains. (c, d, g, h) Comparison between FNN model predicted and GNLSE simulated intensity profiles in temporal and spectral domains, respectively, using the FNN model with lowest NRMSE in each case.

To illustrate how efficient FNN can be, Table 1 compares the computational time between GNLSE simulation time and FNN prediction time for different cases, showing in the format of 'GNLSE simulation time / FNN prediction time'. The time durations are obtained by stopwatch timer function in Matlab. The computation is done on a laptop with 13th Gen IntelI CoreI i5-1335U processor. The FNN model can predict temporal and spectral profiles of a mode-locked fiber laser as fast as a ~5 ms time frame. For the four cases, GNLSE simulations can all converge in 60-80 round trips. Compared with the GNLSE simulation, FNN prediction is 30-60 thousand time faster.

**Table 1. Comparison between consumed time of GNLSE simulation and FNN prediction in different cases.**

| Cases | SSG 11 $m^{-1}$ | SSG 22 $m^{-1}$ | FBW 7 nm | FBW 13 nm |
|---|---|---|---|---|
| Time domain | 181 s/ 4.4 ms | 208 s/ 3.8 ms | 154 s/ 5.1 ms | 156 s/ 3.9 ms |
| Spectral domain | 181 s/ 5.6 s | 208 s/ 3.6 s | 154 s/ 4.8 ms | 156 s/ 4.6 ms |

In conclusion, the excellent regression ability of FNN-trained models is explored to directly predict fiber laser output from cavity parameters with NRMSE below 0.032 and only 5 ms computational time, bypassing the conventional tedious iterative simulation process. Using the FNN-trained models, temporal and spectral profiles of a mode-locked fiber laser operated in the dissipative soliton regime can be accurately predicted. We demonstrated the feasibility and universality of FNN-trained models as a direct mapping tool from cavity parameter to laser output. This can enable the fast, intelligent design of fiber laser cavities with desired output and open the route to autonomous operation maintenance of mode-locked lasers.

**Funding.** Academy of Finland (320165).

**References**

1. G. Genty, L. Salmela, J. M. Dudley, D. Brunner, A. Kokhanovskiy, S. Kobtsev, and S. K. Turitsyn, "Machine learning and applications in ultrafast photonics," Nat. Photonics 15(2), 91–101 (2021).
2. X. Wu, J. Peng, S. Boscolo, Y. Zhang, C. Finot, and H. Zeng, "Intelligent Breathing Soliton Generation in Ultrafast Fiber Lasers," Laser & Photonics Reviews 16(2), 2100191 (2022).
3. J. Girardot, A. Coillet, M. Nafa, F. Billard, E. Hertz, and P. Grelu, "On-demand generation of soliton molecules through evolutionary algorithm optimization," Opt. Lett., OL 47(1), 134–137 (2022).
4. G. Pu, L. Yi, L. Zhang, and W. Hu, "Genetic Algorithm-Based Fast Real-Time Automatic Mode-Locked Fiber Laser," IEEE Photonics Technology Letters 32(1), 7–10 (2020).
5. M. Hary, L. Salmela, P. Ryczkowski, F. Gallazzi, J. M. Dudley, and G. Genty, "Tailored supercontinuum generation using genetic algorithm optimized Fourier domain pulse shaping," Opt. Lett., OL 48(17), 4512–4515 (2023).
6. T. Pinto, U. C. de Moura, F. D. Ros, M. Krstić, J. V. Crnjanski, A. Napoli, D. M. Gvozdić, and D. Zibar, "Optimization of frequency combs spectral-flatness using evolutionary algorithm," Opt. Express, OE 29(15), 23447–23460 (2021).
7. T. Zahavy, A. Dikopoltsev, D. Moss, G. I. Haham, O. Cohen, S. Mannor, and M. Segev, "Deep learning reconstruction of ultrashort pulses," Optica, OPTICA 5(5), 666–673 (2018).
8. A. Kokhanovskiy, A. Bednyakova, E. Kuprikov, A. Ivanenko, M. Dyatlov, D. Lotkov, S. Kobtsev, and S. Turitsyn, "Machine learning-based pulse characterization in figure-eight mode-locked lasers," Opt. Lett., OL 44(13), 3410–3413 (2019).
9. S. Kleinert, A. Tajalli, T. Nagy, and U. Morgner, "Rapid phase retrieval of ultrashort pulses from dispersion scan traces using deep neural networks," Opt. Lett., OL 44(4), 979–982 (2019).
10. M. Närhi, L. Salmela, J. Toivonen, C. Billet, J. M. Dudley, and G. Genty, "Machine learning analysis of extreme events in optical fibre modulation instability," Nat Commun 9(1), 4923 (2018).
11. M. Mabed, F. Meng, L. Salmela, C. Finot, G. Genty, and J. M. Dudley, "Machine learning analysis of instabilities in noise-like pulse lasers," Opt. Express, OE 30(9), 15060–15072 (2022).
12. L. Wang, Z. Ruan, H. Wang, L. Shen, L. Zhang, J. Luo, and J. Wang, "Deep Learning Based Recognition of Different Mode Bases in Ring-Core Fiber," Laser & Photonics Reviews 14(11), 2000249 (2020).
13. U. Teğin, B. Rahmani, E. Kakkava, N. Borhani, C. Moser, and D. Psaltis, "Controlling spatiotemporal nonlinearities in multimode fibers with deep neural networks," APL Photonics 5(3), 030804 (2020).
14. H. Sui, H. Zhu, L. Cheng, B. Luo, S. Taccheo, X. Zou, and L. Yan, "Deep learning based pulse prediction of nonlinear dynamics in fiber optics," Opt. Express, OE 29(26), 44080–44092 (2021).
15. L. Salmela, N. Tsipinakis, A. Foi, C. Billet, J. M. Dudley, and G. Genty, "Predicting ultrafast nonlinear dynamics in fibre optics with a recurrent neural network," Nat Mach Intell 3(4), 344–354 (2021).
16. Y. Fang, H.-B. Han, W.-B. Bo, W. Liu, B.-H. Wang, Y.-Y. Wang, and C.-Q. Dai, "Deep neural network for modeling soliton dynamics in the mode-locked laser," Opt. Lett., OL 48(3), 779–782 (2023).
17. X. Lin, Y. Rivenson, N. T. Yardimci, M. Veli, Y. Luo, M. Jarrahi, and A. Ozcan, "All-optical machine learning using diffractive deep neural networks," Science 361(6406), 1004–1008 (2018).
18. L. G. Wright, T. Onodera, M. M. Stein, T. Wang, D. T. Schachter, Z. Hu, and P. L. McMahon, "Deep physical neural networks trained with backpropagation," Nature 601(7894), 549–555 (2022).
19. T. Wang, S.-Y. Ma, L. G. Wright, T. Onodera, B. C. Richard, and P. L. McMahon, "An optical neural network using less than 1 photon per multiplication," Nat Commun 13(1), 123 (2022).
20. H. A. Haus, J. G. Fujimoto, and E. P. Ippen, "Structures for additive pulse mode locking," J. Opt. Soc. Am. B, JOSAB 8(10), 2068–2076 (1991).
21. Q. Ma and H. Yu, "Artificial Intelligence-Enabled Mode-Locked Fiber Laser: A Review," Nanomanuf Metrol 6(1), 36 (2023).
22. G. Pu, R. Liu, H. Yang, Y. Xu, W. Hu, M. Hu, and L. Yi, "Fast Predicting the Complex Nonlinear Dynamics of Mode-Locked Fiber Laser by a Recurrent Neural Network with Prior Information Feeding," Laser & Photonics Reviews 17(6), 2200363 (2023).
23. J. M. Dudley, G. Genty, and S. Coen, "Supercontinuum generation in photonic crystal fiber," Rev. Mod. Phys. 78(4), 1135–1184 (2006).